\documentclass[twocolumn]{aastex63}

\newcommand{\texp}{t_{\rm exp}}
\newcommand{\diff}{\mathrm{d}}
\newcommand{\avgLBtz}{\left\langle L_{B}\right\rangle}
\newcommand{\feh}{{\rm [Fe/H]}}
\newcommand{\LBtz}{L_{B}}
\newcommand{\mbar}{\overline{m}}
\newcommand{\fdet}{f_{\rm det}}
\newcommand{\ndet}{n_{\rm det}}
\newcommand{\ntot}{n_{\rm tot}}

\newcommand{\Bcut}{B_{\rm cut}}
\newcommand{\Bfs}{B_{5\sigma}}

\newcommand{\Lsqr}{\left\langle L_{B}^2\right\rangle_{L<L_{\rm lim}}}

\received{XXX}
\revised{XXX}
\accepted{XXX}
\submitjournal{AJ}

\shorttitle{Detecting Resolved Populations}
\shortauthors{Lancaster et al.}

\graphicspath{{./}{figures/}}

\begin{document}


\title{{\tt walter}: A Tool for Predicting Resolved Stellar Population Observations\\
with Applications to the Roman Space Telescope}

\correspondingauthor{Lachlan Lancaster}
\email{lachlanl@princeton.edu}

\author[0000-0002-0041-4356]{Lachlan Lancaster}
\affiliation{Department of Astrophysical Sciences, Princeton University, 4 Ivy Lane, 08544, Princeton, NJ, USA}

\author[0000-0003-0256-5446]{Sarah Pearson}\thanks{Hubble Fellow}
\affiliation{Center for Cosmology and Particle Physics, Department of Physics, New York University, 726 Broadway, New York, NY 10003, USA}
\affiliation{Center for Computational Astrophysics, Flatiron Institute, 162 5th Avenue, New York City, NY 10010, USA}

\author[0000-0002-7502-0597]{Benjamin F. Williams}
\affiliation{Department of Astronomy, Box 351580, University of Washington, Seattle, WA 98195, USA}

\author[0000-0001-6244-6727]{Kathryn V. Johnston}
\affiliation{Department of Astronomy, Columbia University, Mail Code 5246, 550 West 120th Street, New York City, NY 10027, USA}
\affiliation{Center for Computational Astrophysics, Flatiron Institute, 162 5th Avenue, New York City, NY 10010, USA}

\author[0000-0003-2539-8206]{Tjitske K. Starkenburg}
\affiliation{Department of Physics \& Astronomy and CIERA, Northwestern University, 1800 Sherman Ave, Evanston, IL 60201, USA}

\author[0000-0002-0332-177X]{Erin Kado-Fong}
\affiliation{Department of Astrophysical Sciences, Princeton University, 4 Ivy Lane, 08544, Princeton, NJ, USA}

\author[0000-0003-0248-5470]{Anil C. Seth}
\affiliation{Physics \& Astronomy, University of Utah, Salt Lake City, UT 84112, USA}

\author[0000-0002-5564-9873]{Eric F. Bell}
\affiliation{Department of Astronomy, University of Michigan, 1085 South University Ave., Ann Arbor, MI 48109-1107, USA}

\begin{abstract}
Studies of resolved stellar populations in the Milky Way and nearby galaxies reveal an amazingly detailed and clear picture of galaxy evolution. Within the Local Group, the ability to probe the stellar populations of small and large galaxies opens up the possibility of exploring key questions such as the nature of dark matter, the detailed formation history of different galaxy components, and the role of accretion in galactic formation. Upcoming wide-field surveys promise to extend this ability to all galaxies within 10~Mpc, drastically increasing our capability to decipher galaxy evolution and enabling statistical studies of galaxies' stellar populations. To facilitate the optimum use of these upcoming capabilities we develop a simple formalism to predict the density of resolved stars for an observation of a stellar population at fixed surface brightness and population parameters. We provide an interface to calculate all quantities of interest to this formalism via a public release of the code: \texttt{walter}. This code enables calculation of (i) the expected number density of detected stars, (ii) the exposure time needed to reach certain population features, such as the horizontal branch, and (iii) an estimate of the crowding limit, among other features. These calculations will be very useful for planning surveys with NASA's upcoming Nancy Grace Roman Space Telescope (Roman, formerly WFIRST), which we use for example calculations throughout this work.
\end{abstract}


\section{Introduction}
\label{sec:intro}
The Nancy Grace Roman Space Telescope (Roman) has the potential to revolutionize the study of stellar populations in nearby galaxies. Roman is a NASA mission currently under construction and scheduled for launch in 2025.  It is a 2.4m telescope designed to cover a $\sim$0.5 degree field of view (FoV) with $\sim$0.1$"$ angular resolution with high sensitivity in 7 broad bands: $F062$, $F087$, $F106$, $F129$, $F158$, and $F184$ \citep{akeson2019}, plus a recently added $K_s$ analog, $F213$\footnote{The naming convention $F{\rm XXX}$ indicates the central wavelength of the filter, i.e. $F062$ and $F158$ correspond to central wavelengths of roughly $0.62\,\mu m$ and $1.58\, \mu m$ respectively.}.  The  large  field  of view and high resolution make Roman ideal for studies of resolved stellar populations in the local universe, which we further motivate in \autoref{subsec:past_science} and \autoref{subsec:future_science} before describing the purposes of this work in \autoref{subsec:this_work}.

\subsection{Recent Resolved Stellar Populations Studies}
\label{subsec:past_science}

Resolved stellar photometry provides a highly sensitive probe for several fundamental astrophysical processes \citep[e.g.,][]{dalcanton2012}. This includes the study of field dwarf galaxies, dwarf satellite galaxies, galaxy stellar halos, and disk formation. In particular, these examples can shed light on the formation of galaxies and the distribution of dark matter. The ages of the stellar populations of field dwarfs probe the earliest epochs of galaxy formation \citep[e.g.,][]{weisz15,fillingham2018,sacchi2021} as well as the epoch of reionization \citep[e.g.,][]{simon21}.

The dwarf satellite mass function's sensitivity to the epoch of reionization makes these low-mass galaxies significant for cosmological constraints \citep[e.g.,][]{bullock2000,graus2016}. The constraining power of galaxy stellar halos has been demonstrated by \citet{bullock05}, who performed a suite of simulations showing the wide array of structures expected for a variety of galaxy formation histories.  Low surface brightness stellar halos provide the strongest known constraints on the accretion histories of galaxies \citep{bullock05,cooper2010,pillepich2014,dsouza2018} and are unique probes of the structure and substructure of the dark matter halos that surround them \citep{johnston99,carlberg09}. 

Observational work has already begun to address some of the cosmic mysteries outlined above. In the Local Group, large new imaging surveys have uncovered dozens of new satellite galaxies and streams around the MW and M31 \citep[e.g.,][and references therein]{martin2013,Drlica15,mcconnachie2018,malhan2018}. This work revealed abundant substructure in halos and discovered dozens of very low mass dwarf galaxies, providing dramatic confirmation of the widely accepted hierarchical galaxy formation paradigm \citep[at least in part;][]{bell08}. This sparked a vigorous theoretical exploration of how the galaxies' assemblies and mass distributions can be accurately measured from the remnants of disrupted satellites \citep{pricewhelan14,sanderson15,GaiaSausage,GaiaEnceladus,Lancaster2pcf,reino20}. The Panchromatic Hubble Andromeda Treasury \citep[PHAT;][]{dalcanton2012} survey observations have revealed detailed insight about the formation and evolution of stars and disk galaxies \citep[e.g,][]{rosenfield12,williams12,lewis15,weisz15,johnson16,williams17} by resolving more than 100 million stars in the disk of M31 \citep{williams2014}.  Outside the Local Group, integrated light and star-counting observations have allowed course mapping of stellar halos as well as the discovery of significant substructures and satellites \citep[e.g.,][]{martinez-delgado2010,vandokkum2014,sand2014,crnojevic2016,carlin2016,mao2021}.
However, due to their faintness and the limited FoV of previous instruments, the resolved populations of stellar halos and low-mass dwarf galaxies have so far only been mapped in detail in a small handful of galaxies \citep[e.g.,][]{newberg2002,majewski2003,calchi2005,ibata2007,mcconnachie2009,crnojevic2016,Bennet19,smercina20}.

\subsection{Potential Future Impact}
\label{subsec:future_science}

These insights have tremendous potential, but with relevant observations available for only a handful of galaxies in a narrow range of environments, definitive empirical conclusions on galaxy assembly writ large and the distribution of dark matter within galaxies remains out of reach. Resolved stellar photometry from Roman will have the potential to discover faint dwarf galaxy satellites in stellar halos \citep[see e.g.,][]{denja19}, adding significantly to the number of galaxies with well-measured satellite mass functions that can be directly compared to simulated galaxy samples \citep[e.g.,][]{elvisI,phatelvis,nihao,satgen,artemis}.  Roman will allow comparisons to be made in new environments for fainter satellite systems beyond the Local Group.  Moreover, the dark matter halos of dwarf galaxies are predicted to be populated by a large number of dark matter clumps \citep{gao04}. Some of these dwarfs may host their own faint satellite galaxies \citep[e.g.,][]{carlin16,carlin20} and streams (see e.g., \citealt{delgado12}, Starkenburg et al., in prep.), making the surroundings of any galaxy (regardless of its mass) a prosperous ground for the discovery of new faint structures \citep{sales2013,wheeler2015,pardy2020}. The statistics of such satellite counts, from dwarf to massive galaxies, can be increased by orders of magnitude with Roman.

Beyond intact satellite galaxies, the physical debris from dwarf galaxies merging with larger halos looks dramatically different if the dwarfs were accreted early/late, on orbits of low/high eccentricity, and if they were of high or low luminosity \citep[e.g.,][]{hendel15}.  The observed properties of substructure can be associated with fundamental physical quantities: the frequency of tidal debris reflects the recent accretion rate; the physical scales and surface brightnesses reflect the mass and luminosity functions of infalling objects; and the morphology reflects the orbits.  Thus, substructure in halos offers a direct constraint on the history and nature of baryonic and dark matter assembly \citep{johnston08}.

In addition to formation processes, tidal debris probes the dark matter properties and distribution around  the Milky Way \citep[e.g.,][]{apw18,koposov10,kuepper15} and external galaxies \citep[e.g.,][]{fardal13, pearson22}. Complexity arises beyond the galaxy accretion history because the morphology and frequency of debris structures can be affected by the three dimensional shape of dark matter halos \citep[certain orbit families can cause ``fanning'' of  thin tidal streams, see][]{pearson15,fardal15,price16,yavetz20}. These ``complications'' leave observable signatures in the morphology of streams alone and hence offer the additional possibilities of measuring the orbit distribution within dark matter halos, and possibly the halos' triaxiality. Additionally, because gaps can form in streams as they interact with dark matter subhalos \citep[e.g.,][]{ibata02,johnston02,yoon11,carlberg12}, particularly thin streams, emerging from globular clusters, are sensitive to low mass subhalos and can be used to indirectly probe the nature of dark matter \citep[e.g.,][]{bovy2017,price18,bonaca2019}. \citet{Pearson19,pearson21} showed that Roman will be able to detect such thin streams in galaxies out to distances beyond 3.5 Mpc. This provides exciting prospects for future searches for gaps in streams orbiting galaxies that do not host molecular clouds, galactic bars or spiral arms which can contaminate the gap signatures from dark matter subhalos \citep[][]{amorisco16,erkal17,pearson2017,banik19}. 
\subsection{This Work}
\label{subsec:this_work}
Capitalizing on these capabilities will require the community to carefully plan observations to efficiently attack the scientific goals of interest. With the aim of providing better tools for planning observations, we develop a general formalism for calculating the number of detected stars per unit sky area in a given observation, and at what point that observation will become crowding limited. This formalism simplifies the estimation of observational sensitivity to surface brightness and stellar population parameters. The formalism itself is also entirely general, and applicable to any observation aiming to resolve a population of stars, not necessarily using Roman.

For the convenience of the future use of this formalism, herein we describe and provide access and installation instructions for {\tt walter} (named in honor of Walter Baade, the first astronomer to resolve M31 in to stars \citep{BaadeM31}), a code to calculate quantities necessary for planning stellar populations surveys. These quantities include the number of stars an observation will detect in a pointing, covering objects from nearby dwarf galaxies that fit entirely within a field to very extended and faint stellar halos of large galaxies. The user can make calculations for populations of a wide range of ages and metallicities.  The code is available on \href{https://github.com/ltlancas/walter}{GitHub}, and only depends on {\tt numpy}, {\tt matplotlib}, {\tt iPython}, and {\tt scipy} \citep{harrisNumpy2020,matplotlib_hunter07,Perez07,scipy}. A faster version of the code used to compute the quantities laid out in this paper also requires {\tt Cython}, though this is not necessary for simpler applications \citep{cython}.

The paper is structured as follows: in \autoref{sec:methods}, we describe our mathematical formalism. In \autoref{sec:examples}, we provide example calculations of the quantities laid out in \autoref{sec:methods}. These calculations are also carried out and described in the accompanying code. In \autoref{sec:obs_test} we give a comparison of the predictions of our code against observational data. Finally, in \autoref{sec:discussion}, we provide concrete examples of how our code could be used to help plan observational campaigns and we give a brief conclusion in \autoref{sec:conclusion}.

\section{Formalism}
\label{sec:methods}

In this Section we explain our approach to calculating the number of stars that are expected to be resolved in a given observation. Additionally, we calculate the point at which an observation would become too crowded to accurately measure the magnitude of all stars that we would like to resolve. To simplify the discussion in this section, we restrict ourselves to single values for the following quantities, which will be the key variables determining an observation:
\begin{itemize}
    \item[$\tau$] - The stellar age of the population.
    
    \item[$\feh$] - The metallicity of the population, defined relative to solar quantities. That is, $\feh = 0$ implies solar iron abundance.
    
    \item[$d$] - The luminosity distance to the population.
    
    \item[$\texp$] - The exposure time of the observation.

    \item[$B$] - The bandpass of the observation, which is a property of the instrument.
    
    \item[$\Sigma_B$] - The surface brightness of the population in the given bandpass.
    
    \item[$\Bfs$] - The 5$\sigma$ limiting apparent magnitude in band $B$ for (isolated) point source detection at an exposure time of 1 hour. This is a property of the telescope/instrument being used.
\end{itemize}
This approach allows us to keep the discussion general and therefore apply the same formalism to any number of different types of resolved stellar population observations, from Milky Way globular clusters to stellar halos of distant galaxies. With this general structure in mind, we proceed by first calculating the expected density of resolved stars per unit sky area for a given observation in \autoref{subsec:ndet}. We then calculate the regime in which an observation will become crowding limited in \autoref{subsec:crowding}. Finally, we discuss some of the caveats of this work in \autoref{subsec:caveats}.

\subsection{Density of Detected Stars}
\label{subsec:ndet}

We calculate the number of resolved/detected stars per unit sky area, $\ndet$, in a given observation defined by the variables specified above: $\tau$, $\feh$, $d$, $\texp$, $B$, $\Sigma_B$, and $\Bfs$ (a parameter of the instrument). We do this by separately calculating (i) the total number-density of stars per unit sky area (detected and undetected), $\ntot$, and (ii) the number fraction of stars that are detected, $\fdet$. We can then calculate $\ndet$ as $\ndet = \ntot \fdet$.

These quantities are intrinsically stochastic, via the stochastic process of star formation. We parameterize this stochasticity by the initial mass function (IMF) of the population, which we denote by $\xi(m)$. We assume that the IMF is fully sampled and use $\xi(m)$ to calculate expectation values for the quantities of interest.  We assume that $\xi(m)$ is normalized so that it integrates to one, this assumption is important for the correctness of our formulae below. The IMF can be thought of as an independent characteristic of the population. In our applications below we make the assumption of a Kroupa IMF \citep{KroupaIMF}, but for now we keep the discussion general. We note that throughout this formalism we refer to the mass $m$ of a star, by which we always mean its \textit{initial mass}, as it is on the Zero-Age Main Sequence (ZAMS).

We first aim to calculate the total number density of stars (detected and undetected) in a given observation, $\ntot$, which depends on all quantities except for the exposure time. 
First, we translate the IMF in to a luminosity function for the population, $\Phi(L)$, using an isochrone, for a given age and metallicity, taken from a stellar evolution code.
We denote the mapping of an initial mass to its luminosity in a given band, $B$, for a given age and metallicity, $\tau$ and $\feh$, as $L_{B,\tau,\feh}(m)$. For brevity of notation we simply write this function as $\LBtz$, where the dependence of the population parameters is implicit.

The fact that this mapping ($m\to \LBtz$) is 
not unique (stars at different masses can have the same luminosity) and therefore non-invertible
is where most of the difficulty of calculating $\ndet$ comes from. This complexity means that we cannot write down a simple expression for the luminosity function, even if our expression for the IMF is quite simple\footnote{Under certain assumptions, simple expressions can be written down for $\Phi(L)$ that are reasonably accurate \citep{Olsen03}.}.

To calculate $\ntot$ we only need $\Sigma_B$, $d$, and the average luminosity of the population in band $B$, $\avgLBtz$, which can be written as:
\begin{equation}
    \label{eq:Lbar}
    \avgLBtz = \int \LBtz(m) \xi(m) \diff m \, .
\end{equation}
Again, the dependence of this quantity on population parameters is implicit. This quantity is the expectation value of the $\LBtz$ map on the space of initial masses. If we first define the distance modulus of the population $\mu \equiv 25 + 5\log_{10}\left(d/1\,{\rm Mpc} \right)$ then we can write $\ntot$ as:
\begin{equation}
    \label{eq:ntot}
    \ntot = \frac{10^{\frac{\Sigma_B - \mu}{-2.5}}}{\avgLBtz} \, .
\end{equation}

Next we wish to calculate the number fraction of the population that we actually observe, $\fdet$. The distance to the population and the exposure time of the observation only contribute to the calculation of $\fdet$ by determining the absolute magnitude at which we can no longer resolve stars, which we write as $\Bcut$ (for an observation in band $B$),
\begin{equation}
    \label{eq:Mlim_def}
    \Bcut = \Bfs - \mu + 1.25 \log_{10} 
    \left( \frac{\texp}{3600 \, {\rm sec}}\right)
\end{equation}
where $\mu$ is again the distance modulus and $\Bfs$ is the apparent magnitude limit for a 1 hour exposure, as defined above. Reference values for $\Bfs$ in each Roman band were obtained from the Roman Wide Field Instrument \href{https://roman.gsfc.nasa.gov/science/WFI_technical.html}{technical specifications webpage}\footnote{\url{https://roman.gsfc.nasa.gov/science/WFI_technical.html}} and are listed in \autoref{tab:band_info}. 

We then define the set of all luminosities brighter than $\Bcut$ as $\mathcal{L}_{\Bcut}$. With these definitions in mind $\fdet$ can be calculated as:
\begin{equation}
    \label{eq:fdetect}
    \fdet = \int_{\LBtz^{-1}(\mathcal{L}_{\Bcut})}\xi(m) \diff m
\end{equation}
where $\LBtz^{-1}(\mathcal{L}_{\Bcut})$ is the set of all initial stellar masses $m$ that are brighter than $\Bcut$ at age $\tau$ and metallicity $\feh$\footnote{I.e. the pre-image of the map $\LBtz$ on the set $\mathcal{L}_{\Bcut}$.}. To be explicit with the dependencies of the $\fdet$ calculation, we write $\fdet(B,\tau,\feh,\Bcut)$, and $\Bcut(d,\texp)$. This is the way that we calculate $\fdet$ as well (in terms of the limiting absolute magnitude) which we then translate in to dependencies on $d$ and $\texp$ using \autoref{eq:Mlim_def} for our calculations.

Putting \autoref{eq:ntot} and \autoref{eq:fdetect} together, we then have the formula for the number of detected stars per unit sky area, $\ndet$, as:
\begin{equation}
    \label{eq:summary}
    \ndet = \frac{10^{\frac{\Sigma_B - \mu}{-2.5}}}{\avgLBtz} 
    \fdet(d,\tau,\feh,\texp) \, ,
\end{equation}
where all of the complexity of stellar populations is now folded in to the quantities $\avgLBtz$ and $\fdet$. 

Below, we review how $\fdet$ and $\avgLBtz$ depend on their various parameters for the Roman bands. However, first, we will include the effects of crowding in this formalism.

\begin{figure}
    \centering
    \includegraphics{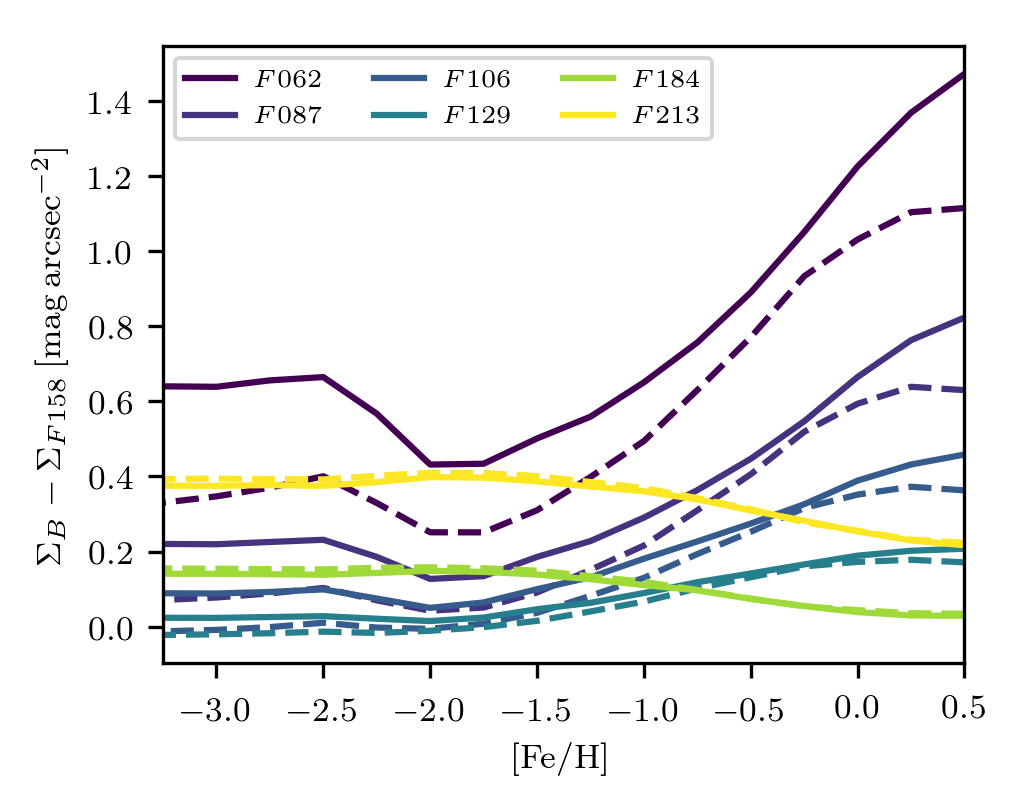}
    \caption{The conversion factors between surface brightness in the $F158$ band and several other bands, indicated by color, as a function of metallicity of the population being observed. We show the conversion factors at an age of $\tau = 2.8$ and $11.2\, {\rm Gyr}$ as dashed and solid lines respectively. These conversion factors can be strong functions of metallicity but are relatively independent of age, especially for the redder bands.}
    \label{fig:Sigma_comp}
\end{figure}

\begin{figure*}
    \centering
    \includegraphics[width=\textwidth]{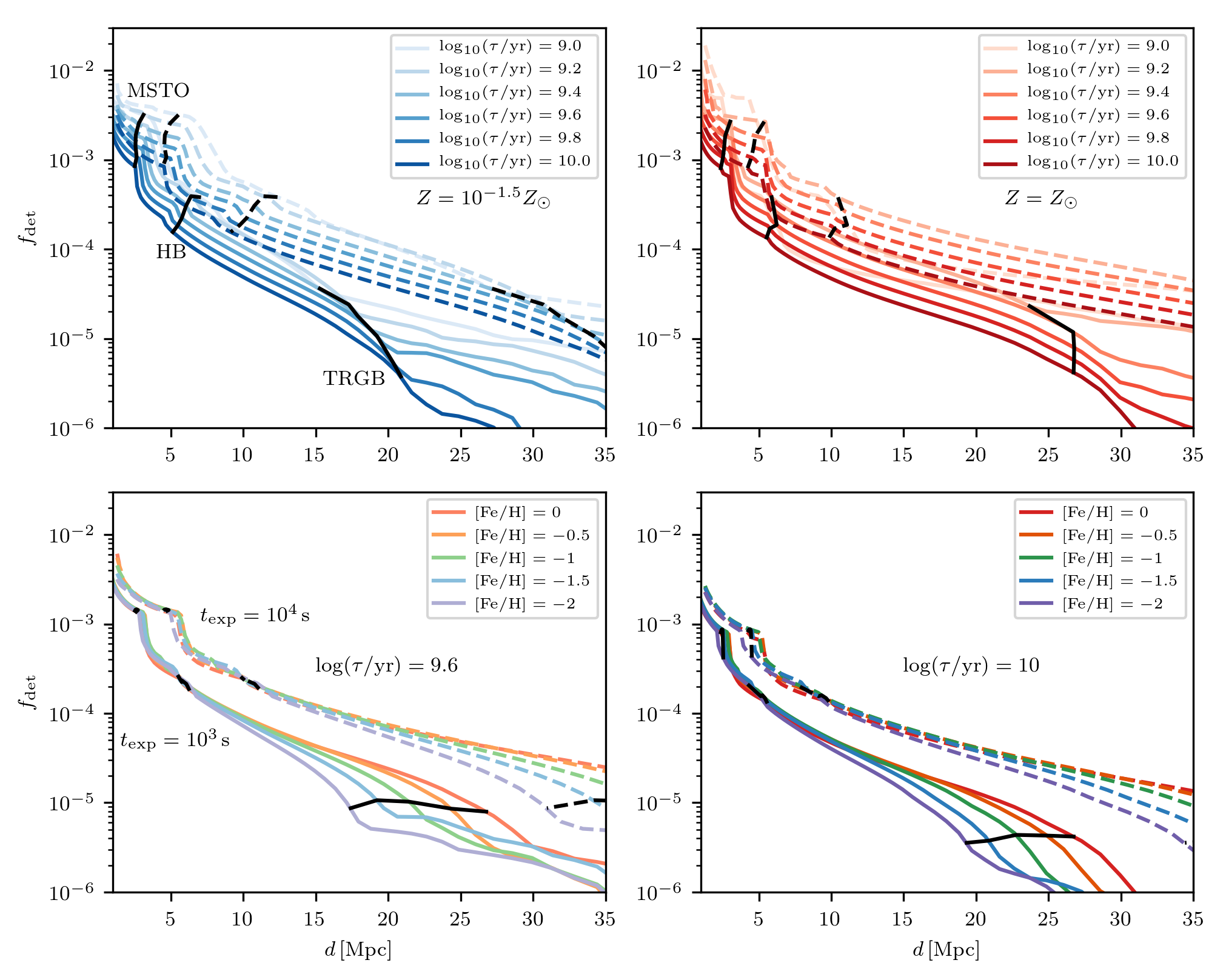}
    \caption{The evolution of the fraction of the stellar population that is detected in the $F158$ band as a function of distance for a $\texp=10^3 \, $seconds (solid lines) and $\texp=10^4 \, $seconds (dashed lines) exposures with Roman and how this varies with the stellar population parameters ($\tau$ in top panels and $\feh$ in bottom panels). For each panel we also indicate the Main Sequence Turn Off (MSTO), Horizontal Branch (HB), and Tip of the Red Giant Branch (TRGB) according to the point at which they fall below the 5$\sigma$ point source detection apparent magnitude. These each correspond to noticeable drops in $\fdet$. Some main take-aways that are apparent in this figure: (i) as a population ages, it dims and, even at fixed surface brightness, we detect a smaller fraction of the populations, (ii) the TRGB can be seen out to $35\, {\rm Mpc}$ for $10^4\, $s exposures and populations of interest to halo studies, and (iii) metallicity mainly has an effect on the observability of the RGB, not other parts of the population.}
    \label{fig:fdet}
\end{figure*}

\begin{figure*}
    \centering
    \includegraphics{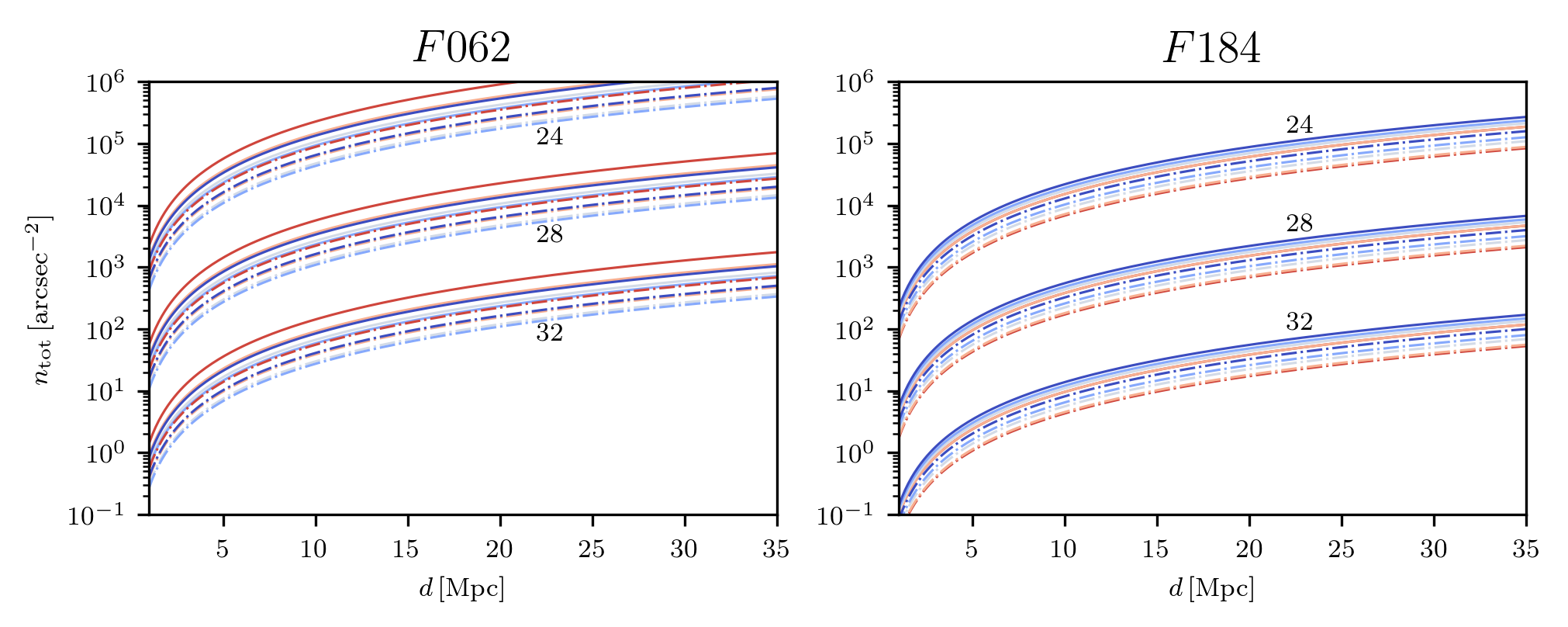}
    \includegraphics{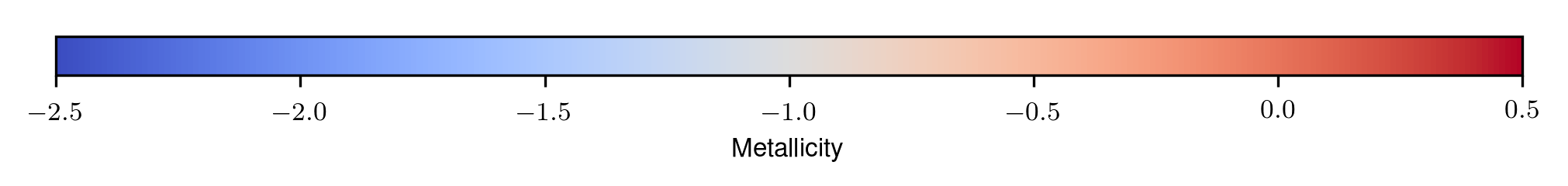}
    \caption{The evolution of the total number density of stars, both detected and undetected, with distance. The panels correspond to two different Roman bands which are indicated in the bottom right corner of each panel. We pick three assumed surface brightnesses, spanning from values typical of the outer disks of galaxies, where crowding can begin to dominate ($24 \, {\rm mag/arcsec}^2$) to the predicted values for faint streams in galaxy halos ($32 \, {\rm mag/arcsec}^2$, see \autoref{fig:halo_obs}). The number density is most strongly dependent on the surface brightness of the population, which is indicated by numbers at the right hand side of each panel in units of AB magnitudes per square arcsecond. The number density is also dependent on the metallicity of the population being considered (which is indicated by the curve colors) and the age of the population, the solid lines indicate an 11 Gyr old population while the dot-dashed lines indicate a 4 Gyr old population. We emphasize the bluest ($R062$) and reddest ($F184$) bands to show the extreme ends of this evolution. Generally, $\ntot\propto d^2$ and increases with age as a population dims (since $\ntot\propto \avgLBtz^{-1}$).}
    \label{fig:ntot}
\end{figure*}

\begin{figure*}
    \centering
    \includegraphics{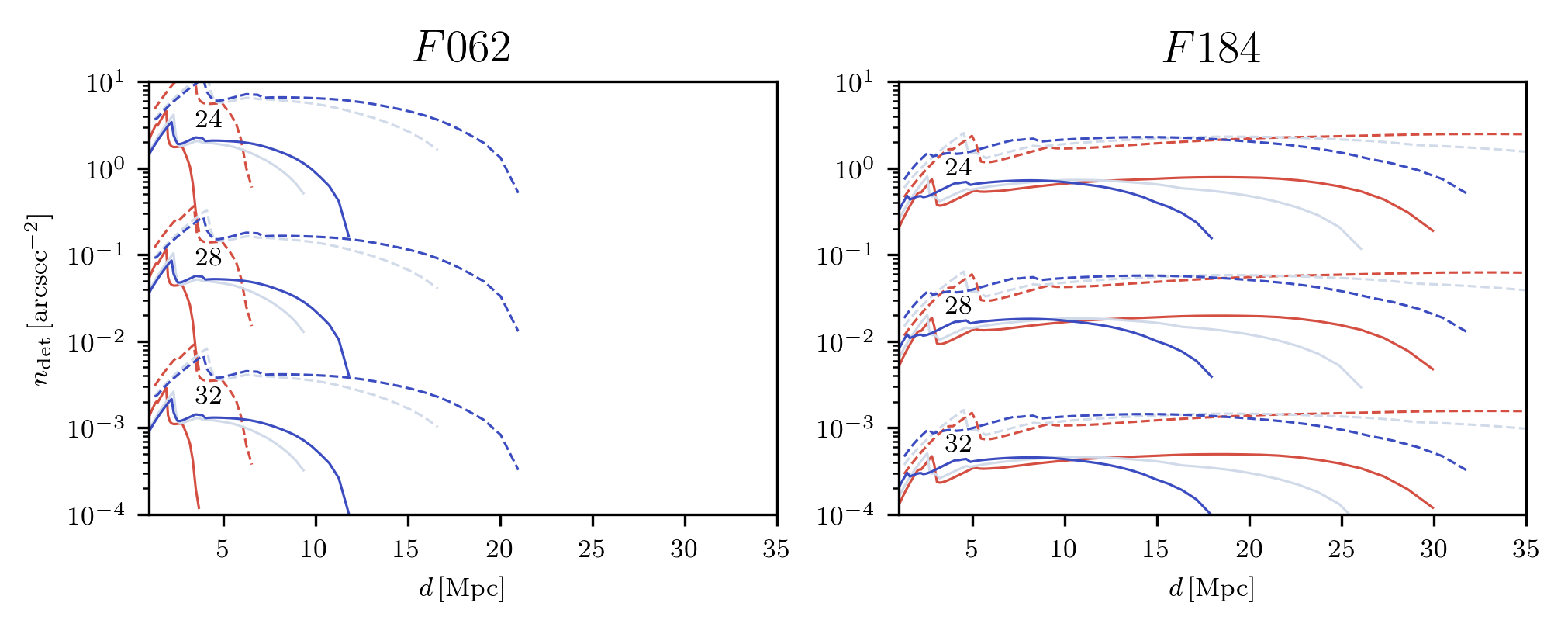}
    \includegraphics{ntot_cbar.png}
    \caption{The evolution of the number density of detected stars with distance for a $10^3\, {\rm s}$ (solid lines) and $10^4\, {\rm s}$ (dashed lines) exposures using the Padova isochrones. We stop plotting the curves once $\fdet<10^{-6}$.  As in Figure \ref{fig:ntot}, the panels correspond to different Roman bands which are indicated in the top right corner of each panel. The number density is most strongly dependent on the surface brightness of the population, which is indicated by numbers at the left hand side of each panel in units of AB magnitudes per square arcsecond. The density is also dependent on the metallicity of the population being considered (which is indicated by the curve colors), and slightly with the age of the population, though we only show a $\tau = 11 \, {\rm Gyr}$ old population here. There are two interesting summary points: (i) The number of detected stars per unit sky area is roughly constant once the Horizontal Branch is no longer detectable as $\ntot \propto d^2$ and $\fdet \propto d^{-2}$ (ii) the redder bands, such as $F184$ here, are extremely helpful in detecting parts of the RGB out to very great distances.}
    \label{fig:ndet}
\end{figure*}

\begin{figure}
    \centering
    \includegraphics{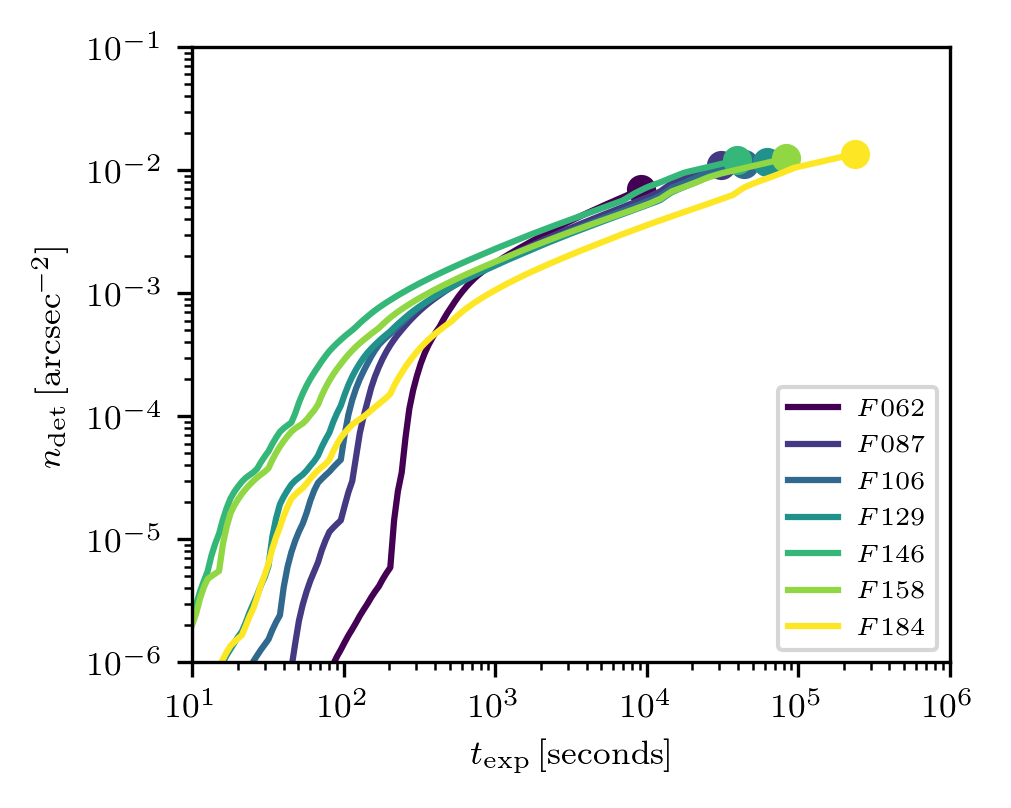}
    \caption{The evolution of the number density of detected stars with exposure time, $\texp$, for a theoretical observation of a population with a surface brightness $\Sigma = 32 \, {\rm mag}\, {\rm arcesec}^{-2}$, an age of $\tau = 11.2\, {\rm Gyr}$, and a metallicity of $\feh = -2$, at a distance of $d = 10\, {\rm Mpc}$. Note, realistically, the same population would not have the same surface brightness in each band. We plot each curve up until the crowding limit is reached for the individual bands, indicated by the circular points. We can see that, for such a low surface brightness, this crowding limit is only reached after a very long integration in most bands. Figures such as this can be very helpful when deciding how to effectively use exposure time. For example, we can see a drastic increase in $\ndet$ for the $F062$ band between 200 and 1000 seconds, with not nearly as much return on invested exposure time afterwards.}
    \label{fig:ndet_texp}
\end{figure}

\subsection{Crowding Limits}
\label{subsec:crowding}

Another important aspect to consider for resolved stellar populations is the crowding limit. This is the surface brightness at which the density of stars on the sky is high enough that it affects the ability to precisely measure the photometry of individual stars. To quantify the effects of the various parameters introduced in the previous section we follow the formalism of \citet{Olsen03}. For a given photometric band $B$, this work calculates the surface brightness at which stars of magnitude $M_{\rm lim}$ can no longer be measured with photometric precision $\sigma_m$ (measured in magnitudes) as:
\begin{equation}
    \label{eq:olsen8}
    \Sigma_m = 2M_{\rm lim}  + \mu - 2.5 \log_{10}
    \left[ \frac{1}{A_{\rm res}} \left( \frac{\sigma_m}{1.086}\right)^2
    \frac{\avgLBtz}{\Lsqr}\right]
\end{equation}
where $\Sigma_m$ is the apparent surface brightness of the population in magnitudes per square arcesecond, $L_{\rm lim}$ is the luminosity corresponding to $M_{\rm lim}$, $\mu$ is the distance modulus, $A_{\rm res}$ is the angular scale of the resolution element or PSF in square arceseconds, and $\Lsqr$ is the expectation value of the square of the luminosity over stars less luminous than $L_{\rm lim}$ defined as:
\begin{equation}
    \label{eq:Lsqr_def}
    \Lsqr \equiv 
    \int_{\LBtz^{-1}(\mathfrak{L}_{L_{\rm lim}})}  
    \LBtz^2(m)\xi (m) \diff m \, ,
\end{equation}
where $\mathfrak{L}_{L_{\rm lim}}$ denotes the set of luminosities dimmer than $L_{\rm lim}$ and $\LBtz^{-1}(\mathfrak{L}_{L_{\rm lim}})$ denotes the set of all initial stellar masses $m$ that are dimmer than $M_{\rm lim}$ at age $\tau$ and metallicity $\feh$\footnote{I.e. the pre-image of the map $\LBtz(m)$ on $\mathfrak{L}_{L_{\rm lim}}$.}.  Note that we have changed the formalism of \citet{Olsen03} to work in terms of the IMF, $\xi(m)$ instead of the luminosity function $\Phi(L)$.

Since we hope to resolve stars up to some limiting magnitude, which we described in the last section as $\Bcut$, for the photometric band $B$, it is natural to set $M_{\rm lim} = \Bcut$ to calculate the crowding limited surface brightness of most interest.

We note that this gives a single surface brightness value at which an observation becomes crowding limited. In reality, a single observation has varying surface brightness over the field of view. This means that some parts of an observation may be crowding limited while others might not.

\subsection{Caveats of our Formalism}
\label{subsec:caveats}

While the formalism laid out above allows the treatment to remain general, it also means that the formalism does not apply to any \textit{real} observation. For example, we assume that the stellar population being observed has a single age and metallicity. In principle, this can be extended by taking a linear combination of the results from single populations in terms of some metallicity/age distribution. However, given our formalism as stated above, this would require knowing the surface brightness of each single-age, single-metallicity component of the distribution. The simpler quantity to work with here would be stellar mass surface density $\Sigma_*$, which could then be straightforwardly and self-consistently translated in to a surface brightness of each component of that age and metallicity. We leave the implementation of this functionality to later updates of the code.

Another side-effect of taking $\Sigma_B$ as our fundamental parameter is that, when performing calculations for multiple bands, it is left up to the user to make sure that the surface brightness in the bands are consistent with one another. As a guide for the case of the Roman bands, we provide the necessary conversions between the surface brightness in the $F158$ band and any other band as a function of metallicity in \autoref{fig:Sigma_comp} for two different ages. We also give numeric values in \autoref{tab:band_info} for a population of age $\tau = 11.2 \, {\rm Gyr}$ and metallicity $[{\rm Fe/H}] = -2$. Though, as we show in \autoref{fig:Sigma_comp}, these conversion factors can be strongly dependent on the metallicity of the population being observed. However, they are less sensitive to the age (at least for not-newly-born populations), especially in the redder bands. The conversion factor between two arbitrary bands $B$ and $C$ is straightforward to calculate from our code as
\begin{equation}
    M_B - M_C = -2.5\log_{10} \left(
    \frac{\left\langle L_B\right\rangle}{\left\langle L_C\right\rangle} \right)
\end{equation}
where $\left\langle L_B\right\rangle$ is given by \autoref{eq:Lbar} for an assumed population age and metallicity. This should be useful for the further application of this code to convert between surface brightnesses observed from ground based observatories to other bands in which one is hoping to resolve the stellar population.

Additionally, our approach is analytic, assuming that the underlying stellar population being observed exactly follows a continuous distribution in mass, $\xi(m)$. This should produce accurate results when the full population being observed fully samples the relevant mass range of the IMF. However, what this mass range is depends on the stellar population properties themselves\footnote{For example, at old stellar population ages, only initial stellar masses $\lesssim 2\, M_{\odot}$ contribute to the luminosity of the population}. Generally this will be true if the inferred total number of stars in the population $N_{\rm tot} = \ntot \Delta \Omega$ (where $\Delta \Omega$ is the area over which the observation is carried out) is large ($\gtrsim 10^4$). $N_{\rm tot}$ can be directly calculated from our code and used as a guide on this front.

There is an additional source of stochasticity when it comes to only detecting the very brightest members of a stellar population. For most reasonable $\xi(m)$, there are only a few stars at the very brightest magnitudes at any age, representing a very small range in mass and thus a very small fraction of the population. This means that while $N_{\rm tot}$ may be large $\fdet$ is very small when only considering the very brightest stars (see \autoref{fig:fdet}) so that the total number of detected stars $N_{\rm det} = N_{\rm tot}\fdet$ is still small. This is especially worrisome since different stellar evolution codes will give slightly different predictions for $\fdet$ at these very low values, meaning that $N_{\rm det}$ can easily go from $10$ to $2$ when using different codes. In general, if the predicted $N_{\rm det}$ of \texttt{walter} is $\lesssim 100$, we advise caution on the strict interpretation of the results.

Taking in to account the assumption of a constant surface brightness is more subtle. In principle, one should be able to break any real observation in to chunks of approximately constant surface brightness, and apply these results individually to each chunk. However, this can be difficult when the surface brightness varies greatly (as in the presence of star clusters) or when the scale on which the surface brightness is constant becomes comparable to the scale on which the IMF is poorly sampled (where our population-averaged formalism would not robustly apply). We do not address this issue here, though it should only be a problem for particularly close-by populations, where the total stellar mass is $\lesssim 10^3 \, M_{\odot}$ over several pixels.

Our formalism does not take in to account crowding by Milky Way foreground stars and background galaxies. As discussed in Appendix A of \citet{Jang20} this can result in a number density of spurious sources of $\sim 3\times 10 ^{-4} \, {\rm arcsec}^{-2}$ for color-magnitude cuts typical of metal-poor RGB stars; somewhat higher background densities, up to $\times 10 ^{-3} \, {\rm arcsec}^{-2}$, are possible if somewhat wider color and magnitude ranges are considered. The observations of \citet{Jang20} had $\texp = 10^3-10^4 \, {\rm secs}$ and were made with the Hubble Space Telescope (HST). As we will show below, this can dominate over the expected number density of detected stars for surface brightness values of interest. This will also effect the crowding and photometric reduction. Therefore, the effect of background sources is an important factor that needs to be taken in to account when applying this formalism to real observations. We leave this for future work.

There are other effects that we do not account for here. For example, the effects of interstellar extinction, which should be small for the infrared wavelengths probed by Roman, could be important for the application of our formalism to other observations. Our formalism also does not address the presence of binary star systems in a population, or binary evolution. This should affect the predictions of this formalism, though likely only by a small factor. Finally, we treat the completeness in detection as going from 100\% at magnitudes brighter than $\Bcut$ to 0\% at magnitudes fainter than $\Bcut$. In reality this will be a smooth transition and could be an important effect in amplifying the number of stars we expect to detect. We leave addressing these issues to future work. Finally, we assume that the distance, $d$, to the stellar population being observed is constant over the observation. While this is usually a safe assumption for extra-galactic observations, it may not be when considering populations in the nearby universe. In principle, this could be addressed in the same way as age and metallicity, using some distribution in distance.

\section{Example Calculations: Resolving Populations with Roman}
\label{sec:examples}

For the convenience of the future use of this formalism, \texttt{walter} is accompanied by example notebooks to walk through the use of the code and how to plot certain quantities. In this section, we explain some of the examples given in the code. These calculations are intended to give the reader an intuitive understanding of the basic dependencies and the order of magnitude values, they are not meant to be exhaustive. For all presented examples, we work under the assumption that our observing instrument is the upcoming Roman Space Telescope. In the code, we point out where changes would need to be made if the code were to be applied more broadly.

To apply the formalism developed in Section \ref{sec:methods} we must choose a mapping from initial stellar mass to luminosity in a given band and for given population parameters, or $\LBtz(m)$, which can be provided by any stellar evolution code. We use two different stellar evolution codes both to give an idea of the uncertainties associated with differences in stellar evolution calculations and because each code has its own benefits and drawbacks. The first code we use is the Modules for Experiments in Stellar Astrophysics (MESA) Isochrones and Stellar Tracks (MIST) code \citep{MIST0,MIST1} and the second is the PARSEC set of isochrones \citep{bressan2012,marigo17}.

Specifically, we download isochrones for stellar populations from metallicity of $0.5$ to $-3.25$ in increments of $0.25\,$dex and ages from $\log_{10} \left(\tau/{\rm yrs}\right) = 8.95$ to $10.1$ in increments of $0.05\,$dex\footnote{We use old stellar ages as we were originally most interested in the stellar halos of galaxies. \textbf{This can be straightforwardly extended to young stellar populations by downloading isochrones for this parameter regime and using the provided code to calculate the quantities of interest.}}. Each isochrone spans a range of initial masses from roughly $0.1M_{\odot}$ to $300M_{\odot}$ and provides a numerical mapping between these initial masses in each photometric band of interest. The photometric bands of interest to us here are the proposed bands for Roman, namely $F062$, $F087$, $F106$, $F129$, $F146$, $F158$, and $F184$. The Padova isochrones also provide predictions for the most recently added filter to the Roman mission, $F213$. Some general information on these bands can be found in \autoref{tab:band_info}. One thing to keep in mind is that all of our formalism is in terms of AB magnitudes whereas the isochrones compute all quantities in Vega magnitudes. For this reason we provide the conversion between the two for each band in \autoref{tab:band_info}.

\begin{deluxetable}{ccccc}
    \tablecaption{Roman Band Specifications \label{tab:band_info}}
    \tablewidth{0pt}
    \tablehead{
    \colhead{Band} &  \colhead{$\Bfs$} 
    &  \colhead{$m_{\rm AB} - m_{\rm Vega}$}
    & \colhead{$\Sigma_B - \Sigma_{F158} $}
    & \colhead{$\Delta \lambda\,[\mu{\rm m}]$}}
    \startdata
    $F062$   & 28.5 & 0.147 & 0.43 & 0.280\\
    $F087$   & 28.2 & 0.485 & 0.13 & 0.217\\
    $F106$   & 28.1 & 0.647 & 0.05 & 0.265\\
    $F129$   & 28.0 & 0.950 & 0.02 & 0.323\\
    $F146$   & 28.4 & 1.012 & 0.05 & 1.464\\
    $F158$   & 28.0 & 1.281 & 0.00 & 0.394\\
    $F184$   & 27.4 & 1.546 & 0.15 & 0.317\\
    $F213$   & 26.2 & 1.819 & 0.40 & 0.350\\
    \enddata
    \tablecomments{Columns are (i) name of the band, (ii) 5$\sigma$ point source detection limiting magnitude for a 1 hour exposure, (iii) conversions between Vega and AB magnitudes, (iv) conversion between surface brightness in $F158$ to a given band for $\tau = 11.2\, {\rm Gyr}$ and $[{\rm Fe/H}] = -2$, and (v) the width of each filter in $\mu$m.}
\end{deluxetable}

To perform all integrals related to Equations \ref{eq:Lbar}, \ref{eq:fdetect}, and \ref{eq:Lsqr_def}, we use the mass samples of the isochrones to integrate over, linearly interpolating between these samples where appropriate cuts must be made. As mentioned earlier, we use a Kroupa IMF with masses limited to being between $0.08 M_{\odot}$ and $120 M_{\odot}$ with $\xi \propto m^{-1.3}$ for $0.08 M_{\odot} \leq m\leq 0.5 M_{\odot}$ and $\xi \propto m^{-2.3}$ for $0.5 M_{\odot} \leq m\leq 120 M_{\odot}$. The code to do all of these integrals is provided with the \href{https://github.com/ltlancas/walter}{GitHub} repository. We also provide an implementation of these integrals in {\tt Cython}, which considerably speeds up their calculation.

\subsection{Calculating $\fdet$}

We begin by illustrating the fraction of all stars in a specified stellar population detectable with a given observation, $\fdet$, and its dependence on the parameters of the observation. As stellar populations generally dim as they age, we expect $\fdet$ to decrease with age, and the dependence on $\texp$ and $d$ are given through the limiting magnitude as stated in Equation \ref{eq:Mlim_def}, however the dependence on metallicity is less clear. To make $\fdet$ more transparent we illustrate its evolution in distance and for several different values of metallicity and age of the observed stellar population in \autoref{fig:fdet} for exposure times of $\texp = 10^3\,$ and $10^4~ {\rm seconds}$ using the MIST isochrones. We additionally indicate several evolutionary stages in Figure \ref{fig:fdet} by associating each stage with the distance at which they are just barely detectable. We do this by associating an absolute magnitude to each stage by evaluating the brightest magnitude within ranges of Equivalent Evolutionary Phase (EEP) provided by the MIST code \citep{MIST0,MIST1}. The ranges in EEP considered for each phase are EEP $\leq 495$ for the Main Sequence Turn Off (MTSO), $630<$ EEP $<640$ for the Horizontal Branch (HB) of Rec Clump (RC), and $560<$ EEP $<580$ for the Tip of the Red Giant Branch (TRGB). These EEP ranges are \textit{not} the default delineations between each phase of evolution given by the MIST team, but we found these choices to more closely match the corresponding drops in $\fdet$ in \autoref{fig:fdet}.

In \autoref{fig:fdet}, we additionally see the expected trend with population age (decreasing $\fdet$ with age) and that the metallicity most noticeably affects the tail-end of the $\fdet$ evolution in distance, indicative of its effect on the luminosity of the RGB. \autoref{fig:fdet} shows that a $10^3\, {\rm seconds}$ exposure with Roman will be able to resolve (detect at $5\sigma$ significance as an isolated source) horizontal branch stars out to a distance of 5~Mpc in the F158 band, and TRGB stars out to about 20 Mpc. \autoref{fig:fdet} also nicely illustrates why $10\, {\rm Mpc}$ is usually thought of as a reference distance for resolved star studies with Roman, as it is past this rough distance that one can no longer easily observe most of the RGB over a large fraction of the parameter space \citep{2015arXiv150303757S}. It is important to note that $10\, {\rm Mpc}$ is not an absolute limit, Roman will be able to see resolved populations much further with longer exposures (subject to crowding limits, discussed further below). Though $10\, {\rm Mpc}$ can be thought of as a reference distance, especially for studies aiming to observe population features like the Horizontal Branch.

\subsection{Calculating $\ntot$}

Now that we have provided a detailed look at the evolution of $\fdet$ with various parameters of the population, we would like to provide some references for the parts of the observation that are independent of the particulars of what is detected. More precisely, for a given surface brightness (which is band dependent) and population, we ask what is the total density of stars that would be expected, both detected and undetected. The real quantity of interest will be a combination of this quantity with $\fdet$, but it is useful to have numbers in mind for the total density of stars. This is given by \autoref{eq:ntot} and its distance evolution is only dependent on the smooth, analytic evolution of the distance modulus so that ($\ntot \propto d^2$ at constant $\Sigma_B$). The exact normalization, however, will depend on the band and population parameters.

To give reference values for these normalizations, we present \autoref{fig:ntot} where we show $\ntot$ for several different values of metallicity (indicated by color), bands ($F062$ and $F184$ as indicated in the bottom right corner of each panel), surface brightness in those bands (indicated by the text adjacent to the curves in each panel), and population ages (indicated by line style) for the Padova isochrones. We see that, as one would expect, the surface brightness of the observed population most strongly determines $\ntot$. We can also see the relative differences between the bluest ($F062$) and second reddest ($F184$) Roman bands, because the populations under consideration are generally dimmer in the bluer bands than the red bands (the average brightness is dominated by the reddest stars) it takes more stars per unit area in the $F062$ band to reach the same surface brightness. This trend is also reflected in the ages, where the younger populations (dot-dashed lines) have fewer stars at fixed surface brightness, since their stars are (on average) brighter.

\subsection{Calculating $\ndet$}

Finally, the quantity of greatest interest to the observer is the density of stars that are detected, described in full by \autoref{eq:summary}. This quantity is essentially the `product' of the quantities laid out in Figures \ref{fig:fdet}, and \ref{fig:ntot}. In \autoref{fig:ndet} we present $\ndet$ in a similar format to that shown in \autoref{fig:ntot}, where the calculations here have been performed using the Padova Isochrones.

We see that the $\ntot \propto d^2$ growth nearly \textbf{compensates} for the drop in $\fdet$ over a large range in distance for a given surface brightness, leading to $\ndet$ being a roughly constant function of distance over quite a large range. We also see the large differences between the $F062$ band and the $F184$ band, where the bluer band ($F062$) detects a much larger number of stars at a fixed surface brightness, and its detection at large distances is much reduced compared to the redder band ($F184$).

Additionally, we see that, in the redder bands, lower metallicity populations generally have larger $\ntot$ at fixed surface brightness (as shown in \autoref{fig:ntot}). They generally have a much larger proportion of these stars below the detection threshold at most exposure times/distances, leading to the lower metallicity populations having slightly larger $\ndet$ at nearby distances, but falling off much earlier than the high metallicity populations. However, this trend is reversed in the bluer band $F062$.

We have thus far shown everything as a function of distance, but the observer will be mainly concerned with picking the correct exposure time for a given observation. In \autoref{fig:ndet_texp}, we give an example of this sort of calculation. We provide a sample of what $\ndet$ would look like for a proposed observation of a $\Sigma = 32\, {\rm mag}\, {\rm arcsec}^{-2}$ (in each band), $\tau = 11.2 \, {\rm Gyr}$, and $\feh = -2$ population at a distance of $d=10 \, {\rm Mpc}$. In \autoref{fig:ndet_texp}, we indicate the point in exposure time at which the observation becomes crowding limited by the points at the end of each curve. These sorts of plots will be extremely useful in making decisions about how to invest exposure time. For example, we see that the most significant return for the observation proposed in \autoref{fig:ndet_texp} is after $\texp = 500 \, {\rm seconds}$ in virtually all bands, after which the slope of each curve becomes significantly less positive. It is also striking that we have any stars at all detected for a 10 seconds exposure, even if it is $\sim 1$ for every million square arcseconds. What we are seeing here is the tail of the asymptotic giant branch (AGB) combined with the fact that these stars are much brighter in the red Roman bands than one would typically expect in comparison to, for example, the optical bands of HST.

\section{Test Against Observations}
\label{sec:obs_test}

In this section we provide a test of our formalism against observations of resolved stellar populations in external galaxies from the ACS Nearby Galaxy Survey Treasury (ANGST) \citep{dalcanton2009,weisz2011a}. To provide the simplest check of the code in its current state we wish to compare against an object of nearly constant surface brightness with as close to a single-metallicity, single-age stellar population as possible. We therefore choose the galaxy FM1 (a.k.a. F6D1) which has a nearly uniformly old ($\sim 11\, {\rm Gyr}$) and metal-poor (${\rm [Fe/H]}\sim - 1.25$) stellar population as inferred from its resolved CMD \citep{weisz2011a}.

\citet{dalcanton2009} report 19,390 stars jointly detected in the F606W and F814W filters of the Hubble Space Telescope's Advanced Camera for Surveys (ACS) with a 50\% completeness limiting magnitude of 28.92 and 27.85 magnitudes, respectively (in the Vega system). Detection in both bands is necessary for removal of contamination from background galaxies \citep[e.g. ][]{muzzin2013}.

In order to make our prediction for the number of stars that should be detected in these observations we take the global surface brightness and size of FM1 from the \citet{karachentsev2013} catalog. Specifically, it is reported that FM1 has an average surface brightness in the $B$ band\footnote{This is the actual Johnson-Cousins $B$ band, not our generic labeling of an arbitrary band $B$ used throughout the rest of the paper.} of $25.8 \, {\rm mag}\, {\rm arcsec}^{-2}$ and an angular diameter of $0.89'$. We also use the distance to the FM1 reported by \citet{dalcanton2009} as $d = 3.4 \, {\rm Mpc}$. In order to calculate the number of expected stars detected in the F814W band we calculate the population-averaged magnitude correction between the F435W (approximately $B$ band) and F814W bands for a stellar population with age $\log_{10}(\tau/{\rm yr}) = 10.05$ and metallicity ${\rm [Fe/H]} = -1.25$ which we find to be $M_{F435W} - M_{F814W} = 1.7$ so that our inferred average surface brightness for FM1 in the $F814W$ band is $24.1 \, {\rm mag}\, {\rm arcsec}^{-2}$.

Using the 50\% completeness limiting magnitude of 27.85 in the $F814W$ band as our $\Bfs$, and the same age and metallicity population used for the correction above we find $\fdet = 4.2 \times 10^{-4}$. With the apparent surface brightness of $\Sigma_{F814W} = 24.1 \, {\rm mag}\, {\rm arcsec}^{-2}$ and $d=3.4 \, {\rm Mpc}$ for the same stellar population we find the total number density of stars (detected and undetected) to be $\ntot = 5884 \, {\rm arcsec}^{-2}$. Taking the angular diameter reported by \citet{karachentsev2013} this implies a total number of detected stars of $10,795$. If we restrict detected stars reported in \citet{dalcanton2009} to this same assumed footprint we find $N_{\rm det} = 5162$. These numbers are not exactly comparable but in the calculation so far we have ignored the effects of extinction and we have assumed 100\% completeness at the 50\% completeness limiting magnitude, both of which would bias us to infer a larger number of stars than are actually detected. If we accept a limiting magnitude that is 0.5 magnitudes smaller (reasonable considering the $0.12 \, {\rm mag}$ estimated extinction in the $F814W$ band; \citealp{2011ApJ...737..103}, and the lack of full completeness) we find a predicted number of detected stars of $5534$. Given the assumptions of the calculation provided here and the fact that the Red Clump lies near the edge of the detection limit for FM1 \citep{dalcanton2009} we believe the agreement of the observed $5162$ stars with the predicted $5534$ stars is a successful test of this framework.

\section{Discussion of Applications}
\label{sec:discussion}

Now that we have produced the tools necessary to determine the sensitivity of observations to a wide range of resolved stellar populations at any distance and surface brightness, we can apply the tools to optimize observing efficiency for observations of nearby galaxies with Roman. Below we provide a few examples of how one might perform such optimizations.  In \autoref{subsec:filter_choices}, we start with optimizing filters for the number of stars detected in a given amount of observing time. In \autoref{subsec:popfeat} we then discuss optimizing observations that wish to detect a given population feature. Lastly, in \autoref{subsec:area} we discuss the example of large halos where we would only cover a fraction of the structures in a single pointing.

\subsection{Filter Choices - One Example}
\label{subsec:filter_choices}

Generally speaking, the choice of filters will depend strongly on the specific science case under consideration or comparability with past measurements (and therefore similarity between filters that have been used in the past). With that in mind, we provide here an example of how one might go about calculating the best filters to use in the case that one is trying to create a color-magnitude diagram (CMD) of an observed population, with no particular interest in any specific part of the CMD. The main considerations in this case would be (i) maximizing the number of stars (probably applicable to many other science goals) and (ii) maximizing the difference in color between the filters used. In this case one will generally have a choice of one redder and one bluer band.

We can then apply our software to determine the optimal exposure time ratios between filters for a given population, and which filters will be best to use within the time constraints. In \autoref{fig:ndet_texp}, we can see that the $F062$ and $F158$ filters reach the highest $\ndet$ in the range $\texp = 10^3-10^4\, {\rm seconds}$\footnote{We exclude the $F146$ band due to its wide wavelength range, which prevents it from providing useful color information.}, which additionally allows for a large color spread. This large color spread is especially important for the creation of CMDs from observations due to the better discerning power on stellar temperatures. Moreover, we can see that at exposure times of roughly $\sim 10^3$ seconds the science return of number of stars per unit exposure time diminishes, indicated by the flattening of each curve. Thus in the case of the observation parameters given in \autoref{fig:ndet_texp}, the most efficient observing plan would be one that exposes $F062$ and $F158$ for $\sim 10^3$ seconds.

While the best filters to use should generally be a function of the population being considered, we would expect the $F062$ and $F158$ filters to generally be good choices for observations aiming to create a CMD, given considerations for depth and color differences. The $F087$ filter could also be a good replacement for the $F062$ filter at shorter exposure times, since it is also a bluer band and reaches larger $\ndet$ at slightly shorter $\texp$. 

\subsection{Detecting a Given Population Feature}
\label{subsec:popfeat}

Another possibility is that the observing program requires that some feature of the stellar population, like the horizontal branch/red clump or TRGB, be detected in at least 4 bands (e.g., to allow for optimal background galaxy separation). This kind of optimization can be determined by producing plots similar to those shown in \autoref{fig:fdet}, which shows the number fraction of the population that is detected as a function of distance at fixed exposure time. In a similar vein, we can also isolate individual population features and calculate the exposure time needed to detect them for a given distance.  We give an example of this in the \texttt{time\_to\_feature} jupyter notebook provided in the code accompanying this paper. As an example, for a population with $\feh= -2$ and $\tau = 10\, {\rm Gyr}$ at 10 Mpc, we can reach one magnitude below the TRGB in 4 bands the fastest if we choose $F106$, $F129$, $F168$, and $F184$, and exposure times of $1777$, $1001$, $451$, $866$ seconds respectively. Given the lack of color information provided by the $F146$ band, we have excluded it from consideration here.

\begin{figure*}
    \centering
    \includegraphics{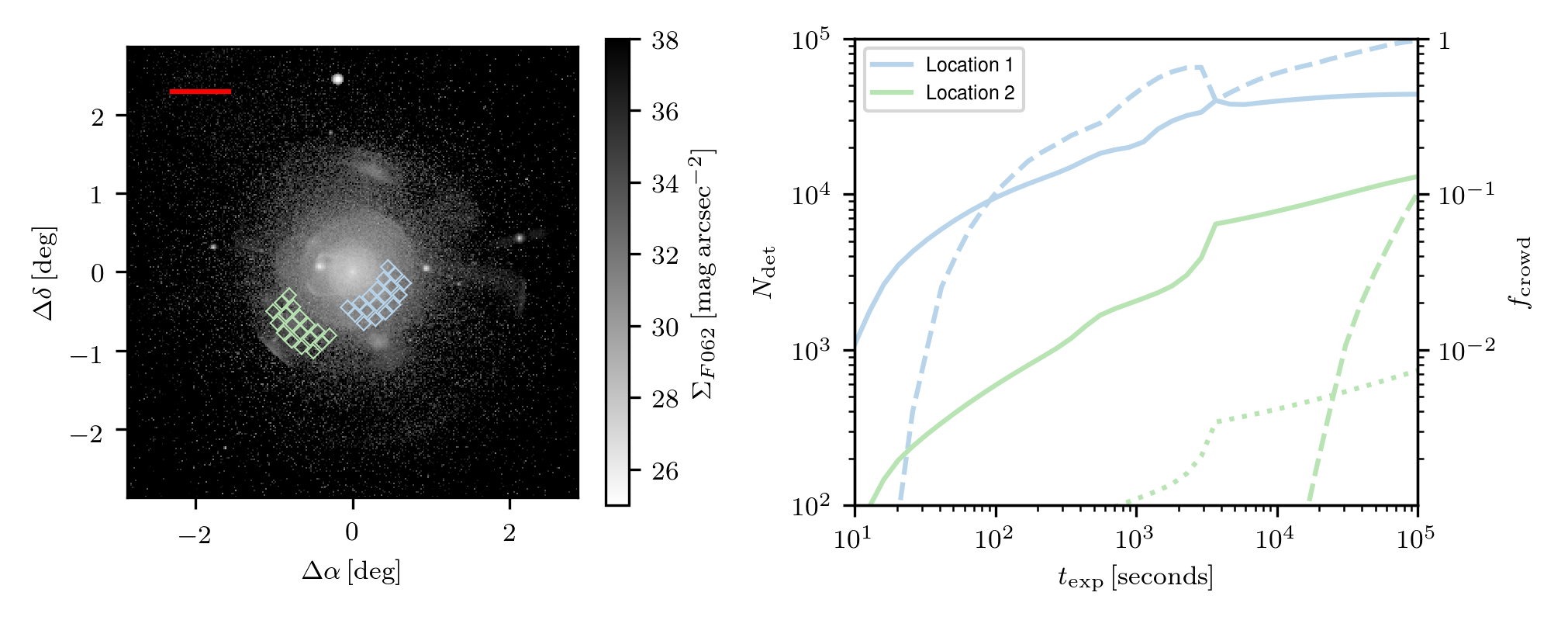}
    \caption{\textit{Left}: A surface brightness map of the Number 2 \citet{bullock05} halo in the $F062$ filter, created as described in the text. The surface brightness is indicated by the color bar. This mock observation assumes that the galaxy is at a distance of 4 Mpc, which sets the scale of the axes in degrees and the size of the RST footprints (shown as light blue and light green tilings, chosen to lie at disjoint locations at different representative galactic radii). We additionally show a red \textbf{bar} that is 50 kpc in length at the upper left, for reference. \textit{Right}: The number of detected stars, $N_{\rm det}$ (solid lines), and the fraction of the observation that is crowded, $f_{\rm crowd}$ (dashed lines), for the two different locations of observations indicated in the left-hand panel. The dotted green line represents the number of detected stars in a giant stream that overlaps the Location 2 field of view.}
    \label{fig:halo_obs}
\end{figure*}

\subsection{Application to Galaxy Halos}
\label{subsec:area}

As a final example, we explore the possibility of mapping a portion of the halo of a nearby galaxy to search for streams.  We start with a model halo from \citet{bullock05}. These simulations consist of stellar tracer particles created by tagging dark matter particles in the simulation and following them as they accrete on to the model halo. We create surface brightness maps from these simulations by projecting the particles along a given axis and binning them on a grid in the other two (non-projected) dimensions. We then calculate the total luminosity in a given band, $B$, from each of these particles by multiplying their masses by $\LBtz/\mbar$, where $\mbar$ is the average stellar mass of the IMF\footnote{This assumes that each stellar tracer particle is massive enough to fully sample the IMF, which is not truly the case in general.}. We then sum the luminosities in each grid cell (pixel), convert this to a magnitude and finally to a surface brightness by adding in the distance modulus and factor accounting for the sky area of each pixel. Since surface brightness is a constant function of distance, a distance is not needed to get the surface brightness map. However, one does need a distance to assign an angular scale to the pixels.

In the left panel of \autoref{fig:halo_obs} we show one of our surface brightness maps in the $F062$ band. Several of the streams have surface brightness $\sim 32 ~{\rm mag/arcsec}^2$, superimposed on a halo background of surface brightness $\sim 37 ~{\rm mag/arcsec}^2$. This map does not include the surface brightness of background galaxies. As we noted in \autoref{subsec:caveats}, our formalism generally ignores the crowding effects of foreground MW stars and background galaxies. The number density of the background galaxies could be as high or higher than the number density of stars detected from a stellar stream, as we will see below.

Subsequently, we perform mock observations of this halo (as carried out in the code provided with this work) by placing a synthetic Roman field of view on the image and predicting the number of stars that would be observed in each RST detector for a given exposure time and two different placements of the detector. These detector placements were chosen to be physically disjoint from one another, in representative areas of the observed halo at different galactic radii. The Location 2 placement was also chosen to lie on top of a giant stream in the simulated halo. For this calculation we have assumed that the halo is at a distance of 4\,Mpc. This calculation assumes a single stellar population, which is not true of the actual simulated halo, though this could be easily extended within \texttt{walter}.

Furthermore, these stellar surface densities allow us to measure the expected amount of crowding, assuming that the crowding limit occurs according to the formalism laid out in \autoref{subsec:crowding}.  We show both the number density of expected stars, and the fraction of the observation that is crowded in the right panel of \autoref{fig:halo_obs}.  

Finally, we can break these numbers down into the number of stars that will be from a feature of interest vs. from the surrounding ambient halo. When interpreting these numbers, it is important to keep in mind that we do not take in to account background sources (as noted in \autoref{subsec:caveats}), which would seriously affect the contrast here.  For example, there is a stream with surface brightness $\sim 34~ {\rm mag/arcsec}^2$ that crosses Location 2, where the ambient background has a surface brightness of $\sim 35 ~{\rm mag/arcsec}^2$.  In a detector that contains the stream, a total of 39,400 stars will be detected in a 1 hour exposure in $F062$.  331 of these stars will belong to the stream. This may not seem like many stars relative to all of the stars collected in the observation, but the stream and stellar halo will be distinct in their spatial and color distribution in ways that are not illustrated by this simple comparison. We anticipate this code being used in future work to provide more quantitative constraints on the ability to detect these tidal features.


\section{Conclusion}
\label{sec:conclusion}

We have developed an open source public software package designed to optimize observing programs aimed at studying resolved stellar structures and resolved stellar populations.  The software package can quickly calculate the number of stars one will detect in an observation if given the population age and metallicity along with the surface brightness, filter and exposure time.  While this number will be useful for many planning purposes, observers must also keep in mind that in addition to the detected stars, which our software can predict, there will be background galaxies that will need to be filtered and taken into account for detection of features.  The code is available on \href{https://https://github.com/ltlancas/walter}{GitHub}, has minimal dependencies, and is laid out with specific applications to the Roman Space Telescope.

We have shown how this software can be used to optimize observing efficiency for a few example programs that one might consider, including determining the best filter choices for a particular science case, detecting dwarf galaxies, and searching for stellar streams to a specific surface brightness limit.  There will likely be many other cases for which this package will be useful, and we hope that it encourages the community to get involved in planning potential General Astrophysics Observations with the Nancy Grace Roman Space Telescope.

\begin{acknowledgements}

LL thanks David Sand, Johnny Greco, Scott Carlsten, and Rachael Beaton for helpful discussions. The authors thank Leo Girardi for providing access to the low metallicity and $F213$ predictions for the Padova isochrones. Support for SP was provided by NASA through the NASA Hubble Fellowship grant \#HST-HF2-51466.001-A awarded by the Space Telescope Science Institute, which is operated by the Association of Universities for Research in Astronomy, Incorporated, under NASA contract NAS5-26555. The Flatiron Institute is supported by the Simons Foundation. 
Support for BFW, KVJ, and AS was provided through NASA contract NNG16PJ28C as part of the Roman Science Investigation Team funded through ROSES call NNH15ZDA001N-WFIRST.

\end{acknowledgements}

\software{
{\tt scipy} \citep{scipy},
{\tt numpy} \citep{harrisNumpy2020}, 
{\tt IPython} \citep{Perez07}, 
{\tt matplotlib} \citep{matplotlib_hunter07},
{\tt Cython} \citep{cython},
}




\bibliography{bibliography}{}
\bibliographystyle{aasjournal}

\end{document}